\documentclass[aip,jcp,twocolumn,showpacs,superscriptaddress,amsmath,amssymb,floatfix,10pt]{revtex4-1}  
\usepackage{graphicx}  
\usepackage{dcolumn}   
\usepackage{bm}        
\usepackage{amssymb}   
\usepackage{framed}
\usepackage{amsmath}

\begin{document}

\widetext
\title{Performance of Range Separated Hybrids: Study within BECKE88 family and Semilocal Exchange Hole based 
Range Separated Hybrid}

\author{Subrata Jana}
\email{subrata.jana@niser.ac.in}
\affiliation{School of Physical Sciences, National Institute of Science Education and Research,
Bhubaneswar 752050, Homi Bhava National Institute, India}
\author{Bikash Patra}
\email{bikash.patra@niser.ac.in}
\affiliation{School of Physical Sciences, National Institute of Science Education and Research,
Bhubaneswar 752050, Homi Bhava National Institute, India}
\author{Hemanadhan Myneni}
\email{hemanadhiitk@gmail.com}
\affiliation{Department of Physics and Astronomy, University of Delaware, Newark, Delaware 19716, USA}
\author{Prasanjit Samal}
\email{psamal@niser.ac.in}
\affiliation{School of Physical Sciences, National Institute of Science Education and Research,
Bhubaneswar 752050, Homi Bhava National Institute, India}

\thispagestyle{empty}

\date{\today}

\begin{abstract}
A long range corrected range separated hybrid functional is developed based on the density matrix expansion (DME) 
based semilocal exchange hole with Lee-Yang-Parr (LYP) correlation. An extensive study involving the proposed
range separated hybrid for thermodynamic as well as properties related to the fractional occupation number is compared with 
different BECKE88 family semilocal, hybrid and range separated hybrids. It has been observed that using Kohn-Sham 
kinetic energy dependent exchange hole several properties related to the fractional occupation number can be improved 
without hindering the thermochemical accuracy. The newly constructed range separated hybrid accurately describe 
the hydrogen and non-hydrogen reaction barrier heights. The present range separated functional has been 
constructed using full semilocal meta-GGA type exchange hole having exact properties related to exchange hole therefore,
it has a strong physical basis. 
\end{abstract}

\maketitle
\section{Introduction}
Density functional theory (DFT)~\cite{ks65,Parr89,RMDreizler90} is {\it de-facto} the standard 
many-body formalism to calculate the electronic and structural properties of atoms, molecules, and solids from small 
to large scale. It is an exact theory only if the exchange-correlation (XC) energy functional, which contains all the 
crucial many body effects are known exactly. But, practically, Kohn-Sham (KS) formalism uses approximate XC functional 
because the exact form of XC functionals is unknown. In DFT there is mainly two class of approximations that have 
been used widely. One is semilocal formalism and other is the hybrid functional theory~\cite{PW86,B88,LYP88,BR89,
B3PW91,B3LYP,PBE96,VSXC98,HCTH,PBE0,HSE03,AE05,MO6L,TPSS03,revTPSS,PBEsol,SCAN15,Kaup14,Tao-Mo16,con1,con2,con3,con4}. 
All these approximations are classified through the Jacob's Ladder approximations, which arranges all the XC according to 
their complexity and accuracy. The first three rungs of the Jacob's ladder are recognized as the local density approximation 
(LDA)~\cite{VWN80,PW92}, generalized gradient approximation (GGA)~\cite{PBE96,PW91,ZWu06} and meta-generalized gradient
approximation (meta-GGA)~\cite{TPSS03,MO6L,SCAN15,Tao-Mo16}. The GGAs and meta-GGAs are extensively used due to their low 
computational cost and accuracy in chemistry and condensed matter physics.~\cite{CJCramer09,Quest12,Yangreview,
Becke14}. All these semilocal functional are computationally very cheap because they involve density ($\rho$), 
gradient of density ($\nabla\rho$) and Kohn-Sham kinetic energy density ($\tau=\frac{1}{2}\sum_i^{occ}|\nabla\psi_i|^2$) 
as their ingredients. Semilocal functionals achieve remarkable accuracy in describing atomization energies~\cite{VNS03,
PHao13,LGoerigk10,LGoerigk11}, surface properties~\cite{YMo16} of solids, bulk modulus~\cite{YMo16}, lattice constants
~\cite{Csonka09,PBlaha09,FTran16}, bond lengths~\cite{CAdamo10,YMo16}, cohesive energy~\cite{VNS04}. But, due to the 
absence of desired non-locality~\cite{PSTS08,pezb98} of the semilocal approximations of exchange hole they drastically 
failed in few cases especially in describing excited state properties such as band gap, excitation energies, reaction 
barrier heights~\cite{vsp07}, Rydberg excitation, the stability of anions and dissociation curve~\cite{mcy06,rpcvs06}. 

The source of errors in  describing all the above phenomena accurately within semilocal formalism can be understood as 
the lack of many electron self-interaction (MESI)~\cite{vsp07,mcy06,rpcvs06,pz81,prp14,hkkk12,phl84,rpcvs07,pplb82,rmjb17}. 
It has been observed that semilocal functional accurately describe the total energy at integer particle numbers but due to 
the lack of self-interaction error (SIE) they deviate from exact curve at fractional occupation number~
\cite{vsp07,rpcvs06,rpcvs07,rmjb17}. Unlike Hartree-Fock (HF), in semilocal functionals, the exchange energy only partially 
get canceled by the coulomb interaction for one electron. However, it is observed that the problem of MESI  can be 
abated by mixing part of HF with semilocal functionals. All these are known as the hybrids functional~\cite{B3PW91,B3LYP,
PBE0,HSE03,rmjb17,hse06,VNS03,JJaramillo03,camb3lyp,mcy06,lcwpbe,hs10,jks08,whc,tpssrs}, which mix the HF either in globally 
(hybrids)~\cite{B3LYP,PBE0,VNS03} or only in some portion either in long range or in short range (range separated hybrids)
~\cite{hse06,camb3lyp,mcy06,lcwpbe,rmjb17,tpssrs}. The essence of mixing HF with the semilocal was first realized by Becke~
\cite{B3LYP}. Among the hybrids the most popular one B3LYP~\cite{B3LYP}, PBE0~\cite{PBE0}, LC-$\omega$PBE~\cite{lcwpbe}, 
HSE06~\cite{hse06}. All the range separated functions use HF in the long range to improve the asymptotic correction of potential. 
Only, exceptional is HSE06, which uses HF in its short range. The HSE06 has been designed from solid state properties point of 
view as solids have no asymptotic tail and, it has been observed that using HF in the short range reduces the computational cost 
of calculation. The HSE06 is very popular in describing solid state band gap. Whereas, LC-$\omega$PBE very popular in predicting 
reaction barrier height. Among Others functionals, the CAM-B3LYP are long range corrected and remarkably accurate in describing 
dissociation curve and reaction barrier heights over the B3LYP hybrid. The CAM-B3LYP~\cite{camb3lyp} are obtained from the thermochemical 
point of view, which uses B88 exchange with both Lee-Yang-Parr (LYP)~\cite{LYP88} and Vosko-Wilk-Nusair (VWN)~\cite{VWN80} correlation. 
Later, it has been observed that further modifications over CAM-B3LYP can be done by tuning the parameters so that the fractional 
energy curve may be made close to that of exact one~\cite{mcy06}. Removing MESI improve many properties especially dissociation limit 
and reaction barrier. Not only that, very recently, further modification over CAM-B3LYP has been attempted~\cite{rmjb17} using Bartlett's 
ionization potential (IP) theorem~\cite{bert1,bert2,bert3}. One way or other there have been several ways to fix the parameter involved 
in the range separately functionals. The modified functionals obtained by using Bartlett's ionization potential (IP) theorem coupled with 
CAM-B3LYP are known as CAM-QTP00 and CAM-QTP01, which are the possible modifications within the CAM-B3LYP functionals.

The range separated functional is constructed by dividing the Coulomb operator into short range and long range part by combining with error 
function. The semilocal short range part of exchange are constructed by combining error function with the semilocal exchange hole.
In the range separation functionals, the range separation is involved only in the exchange part. Whereas, there is no range separation 
involved in correlation part because correlation is essential for both short range and long range. All B88 family range separated functionals 
(LC-BLYP,CAM-B3LYP, rCAM-B3LYP, CAM-QTP00 and CAM-QTP01) are constructed using LDA exchange hole where the inhomogeneity is taken care by 
passing it into the Thomas-Fermi wave vector as prescribed in ~\cite{ldars,ityh01}. Beyond te LDA exchange hole Tao-Mo~\cite{Tao-Mo16} recently 
developed density matrix expansion (DME) based semilocal exchange hole, which satisfy some exact constraint like (i) recovering uniform density limit, (ii)correct small $u$ expansion and (iii) large $u$ diverges. All these properties coupled 
with the inhomogeneity included in the $k$ vector through the normalization of exchange hole establish its superiority over 
all other previously proposed exchange hole. In this paper, the DME based exchange hole has been used as a range-separated 
hybrid by suitably mixing HF exchange with the semilocal meta-GGA type DME exchange hole. The present attempt  is to 
study the thermodynamic properties, the role of fractional occupation number on an atom, dissociation curve and fractional 
occupation number on dissociation limit of our DME based range separated functional. All those properties are compared  
with B88 family semilocal (BLYP), hybrid (B3LYP) and range separated hybrid functionals (LC-BLYP, CAM-B3LYP, rCAM-B3LYP, CAM-QTP00 
and CAM-QTP01). The main reason to choose only B88 family because our DME based range separated functional also use LYP 
correlation and exchange part is derived from spherical averaged exchange hole instead of reversed engineered of system averaged 
exchange hole~\cite{tpsshole,Lucian13}. System averaged exchange hole was used in LC-$\omega$PBE and HSE06 hybrids. Also, we observed 
that using TM proposed  meta-GGA correlation~\cite{Tao-Mo16} with our DME exchange hole based range separated functional not producing 
satisfactory results. Our present paper is organized as follows: in the next section, we will briefly discuss the range 
separated functionals beyond the B3LYP hybrid functional,  then we will discuss the construction of our functional. After that extensive 
study of our function for thermodynamic properties and properties related to the fractional occupation numbers will be studied. 
Lastly, a comparison between hybrids and range-separated hybrids are discussed.  

\section{Range Separated Functionals Beyond B3LYP Hybrid}
The KS DFT ground state total energy is a unique functional of density and given by,
\begin{equation}
\begin{split}
 E^{DFT}[\rho]=\sum_{i}^{\sigma,occ}n_{i,\sigma}\langle\psi_{i,\sigma}|-\frac{1}{2}\nabla^2 |\psi_{i,\sigma}\rangle
 +\int~\rho(\mathbf{r})v_{ext}(\mathbf{r})d\mathbf{r}\\
 +J[\rho]+E_{xc}[\rho_\sigma],
 \label{eq1}
 \end{split}
\end{equation}
where
\begin{equation}
J[\rho]=\frac{1}{2}\int\int\frac{\rho(\mathbf{r})\rho(\mathbf{r'})}{|\mathbf{r}-\mathbf{r}'|}d\mathbf{r}d\mathbf{r'}
\label{eq2}
\end{equation}
is the Hartree energy and  $E_{xc}[\rho_\sigma]$ be the exchange-correlation functional, which has to be approximated within or 
beyond semilocal form. The total density $\rho(\mathbf{r})$ is the sum of spin polarized densities so that,
\begin{eqnarray}
 \rho(\mathbf{r})=\sum_{\sigma=\uparrow~or~\downarrow}\rho_\sigma(\mathbf{r})&=&\sum_{i}^{\sigma,occ}
 n_{i,\sigma}|\psi_{i,\sigma}|^2\nonumber\\
 &=&\sum_{i}^{occ}|\psi_{i,\uparrow}(\mathbf{r})|^2+\sum_{i}^{occ}|\psi_{i,\downarrow}(\mathbf{r})|^2
 \nonumber\\
 &=&\rho_{\uparrow}(\mathbf{r}))+\rho_{\downarrow}(\mathbf{r})
\label{eq3}
 \end{eqnarray}
Upon taking the functional derivative of Eq.(\ref{eq1}), one arrives at the single particle KS equation,
\begin{equation}
 [-\frac{1}{2}\nabla^2+v^{KS}_\sigma(\mathbf{r})]\psi_{i,\sigma}=\epsilon_{i,\sigma}\psi_{i,\sigma},
 \label{eq4}
\end{equation}
where
\begin{equation}
 v^{KS}_\sigma(\mathbf{r})=v_{ext}(\mathbf{r})+\frac{1}{2}\sum_\sigma\int\int\frac{\rho_\sigma(\mathbf{r})}
 {|\mathbf{r}-\mathbf{r}'|}d\mathbf{r}+v_{xc,\sigma}(\mathbf{r})
 \label{eq5}
\end{equation}
and $\epsilon_{i,\sigma}$ is the KS orbital energy (or, KS eigenvalues). The main challenge of KS DFT is to design the XC 
functionals. Semilocal functionals are designed to compensate both the accuracy and computational cost. Beyond LDA (which 
is the lowest rung of Jacob's Ladder), GGA and meta-GGAs are also proposed with increased accuracy. On the $4^{th}$ rung 
of Jacob's Ladder, there are Hybrid functionals, which solves many density functionals problems that are not achievable 
within semilocal density functional approximations (DFA). Hybrid functionals are proposed by mixing some portion of density 
functional approximations (DFA) with non-local HF. One of the most popular Hybrid functional is B3LYP. It accurately describes 
the thermochemical properties. Hybrid functionals use HF exchange without considering any range separation. In spite of its grand 
success, it is failed to accurately describe dissociation curve, reaction barriers heights and properties associated with 
fractional occupation numbers. It has been shown that B3LYP is over delocalized in case of fractional occupation number and 
wrongly predicts the dissociation curve. Therefore, it is a great challenge to describe thermochemical properties as well as 
describe the problems related to fractional occupation numbers on an equal footing. The search for better functional than B3LYP is therefore always been an intriguing research 
field. The factional charge perspective can be understood from Janak's theorem~\cite{janak}, 
\begin{equation}
\frac{\partial E}{\partial n_{i,\sigma}} = \epsilon_{i,\sigma}
\label{eq6}
\end{equation}
It has been proved from the generalized Kohn-Sham formalism (GKS)~\cite{cmy08} that for a fixed configuration (potential), 
the energy is stationary w.r.t the potential, only the frontier level occupation $n_{i=f}$ able to change ($\delta n_{i=f}=q$). 
Hence, 
\begin{equation}
 \frac{\partial E_v}{\partial N}=\Big(\frac{\partial E_v}{\partial n_{i=f}}\Big)_v=\epsilon_{i=f,\sigma},
 \label{eq7}
\end{equation}
where $n_f=n_{LUMO}$ if we fill up the highest unoccupied orbital ($q>0$) and $n_f=n_{HOMO}$ if we remove particles from the highest 
occupied orbital ($q<0$).  The seminal work of Perdew et. al.~\cite{pplb82} shows that energy is piecewise 
linear in between addition of removing the fractional number of electrons. If one adds fraction of particle $q$ to the ground 
state then energy becomes,
\begin{equation}
 E(N_0+q) = (1-q)E(N_0)+q E(N_0+1)~~~~0\leq q\leq 1
 \label{eq8}
\end{equation}
and 
\begin{equation}
 E(N_0+q) = (1+q)E(N_0)-q E(N_0-1)~~~~-1\leq q\leq 0
 \label{eq9}
\end{equation}
It has been observed that all semilocal density functional approximations (DFA) failed to achieve the piecewise linearity and predicts wrong energy 
for fractional occupation number and are highly delocalize. Whereas, HF is highly localized due to the lack of correlation. This affects 
thermochemical phenomena like barrier heights, dissociation limit, Rydberg and charge transfer phenomena. Deviation 
of linearity for fractional charge also observed in the case of B3LYP hybrid, though it is a mixer of DFA (highly 
delocalized) and HF (highly localized). On the next level of B3LYP hybrid functional, it is proposed to improve the 
properties related to fractional occupation number by using the long range corrected hybrid functional like CAM-B3LYP, 
LC-BLYP, rCAM-B3LYP, CAM-QTP00, CAM-QTP01. The general form of long range corrected B3LYP family hybrid is given by,
\begin{equation}
\begin{split}
 E_{xc}^{B88-RS-FAMILY}=\alpha E_x^{HF}+(1-\alpha-\beta)E_x^{B88}+\beta E_x^{HF}\\+\gamma E_x^{LYP}
 +(1-\gamma)E_c^{VWN5}+\delta\Delta E_x^{B88},
 \label{eq10}
 \end{split}
\end{equation}
The range separated parameters $\mu$ along with $\alpha$, $\beta$ enter into the error function in the following way,
\begin{equation}
 \frac{1}{r_{12}}=\underbrace{\frac{\alpha+\beta erf(\mu r_{12})}{r_{12}}}_{LR}
 +\underbrace{\frac{1-[\alpha+\beta erf(\mu r_{12})]}{r_{12}}}_{SR}
\label{eq11}
 \end{equation}
In B88 family range separated hybrids HF is used in LR part whereas DFA is used in SR part. The range separated parameter 
$\mu$, $\alpha$ and $\beta$ control the amount of short-range DFA and long-range HF Exchange. As $r_{12}\rightarrow 0$, 
the HF exchange fraction is $\alpha$, while the DFA exchange fraction is $(1 - \alpha)$. As  $r_{12} \rightarrow \infty$, 
the HF exchange fraction approaches $\alpha + \beta$ and the DFA exchange fraction approaches $(1 - \alpha - \beta)$. 

\begin{figure*}[h]
\includegraphics[width=.40\linewidth]{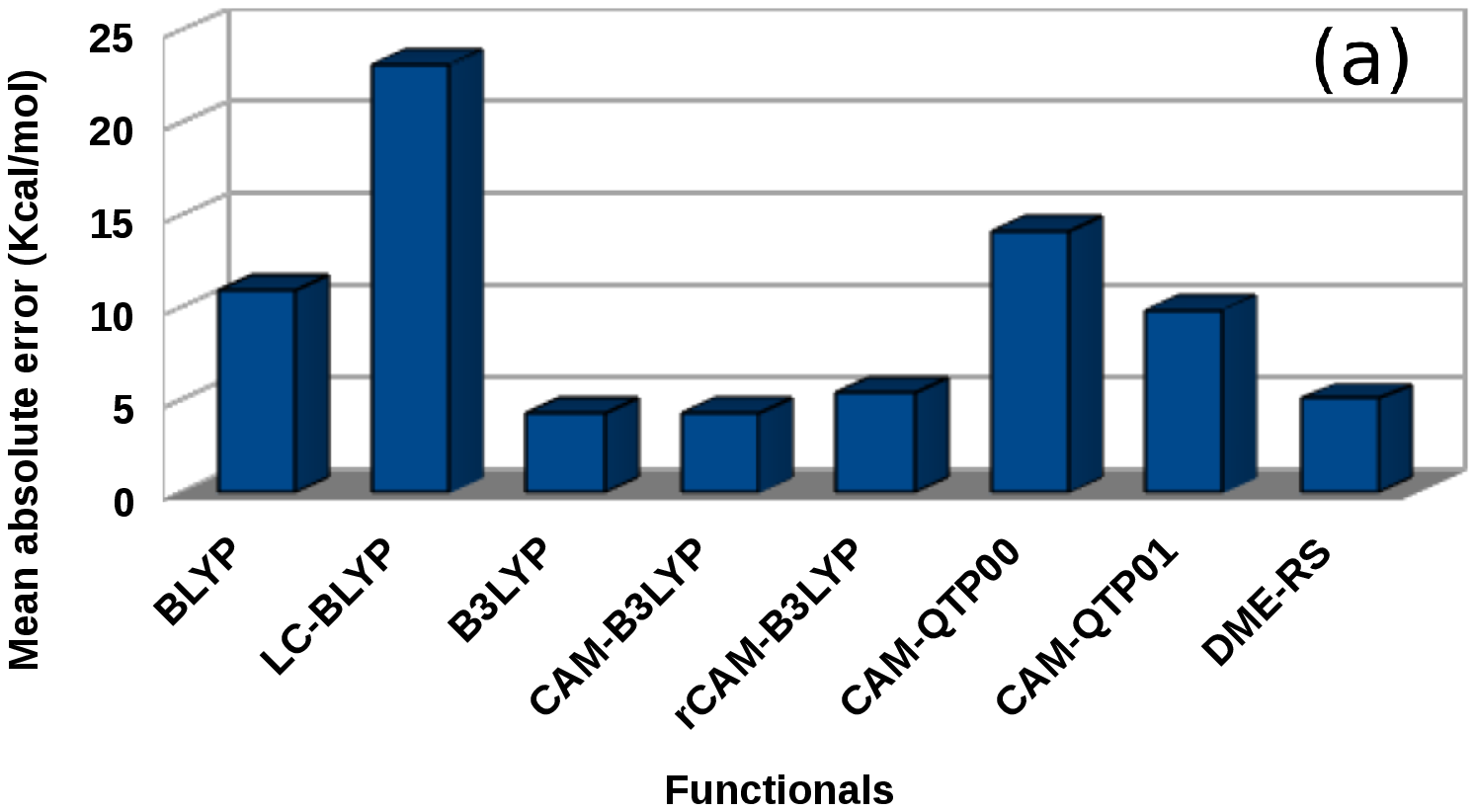}
\includegraphics[width=.40\linewidth]{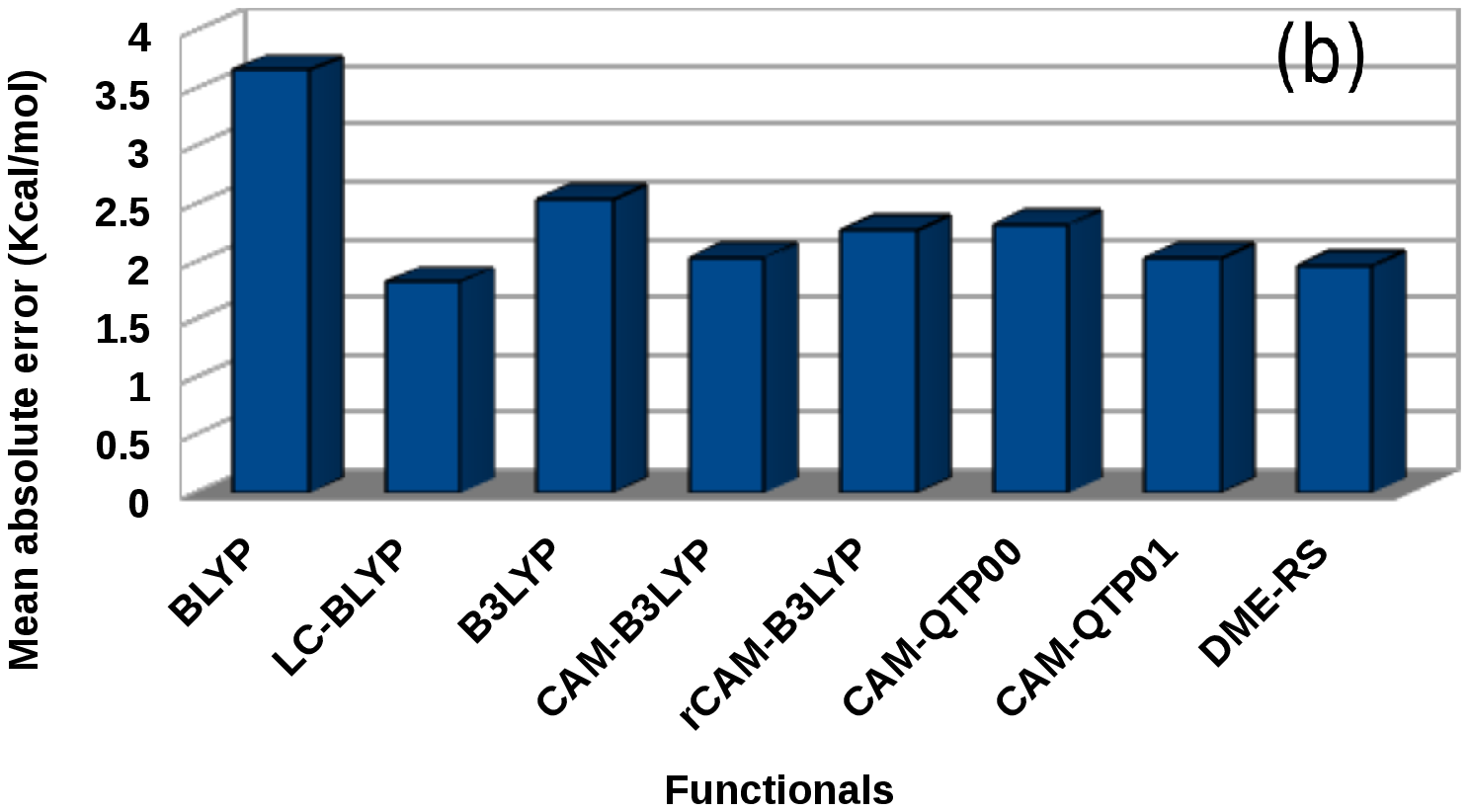}
\includegraphics[width=.40\linewidth]{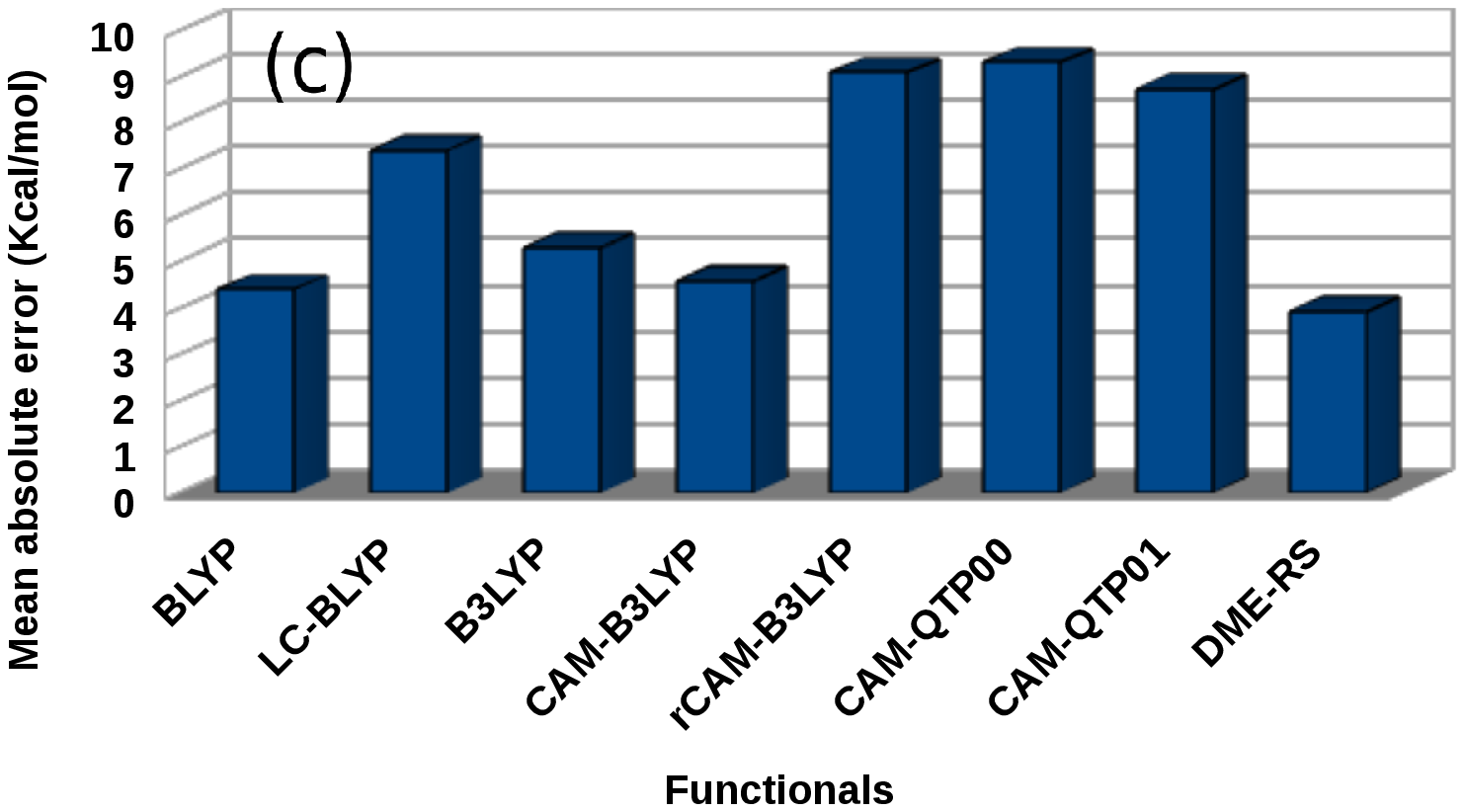}
\includegraphics[width=.40\linewidth]{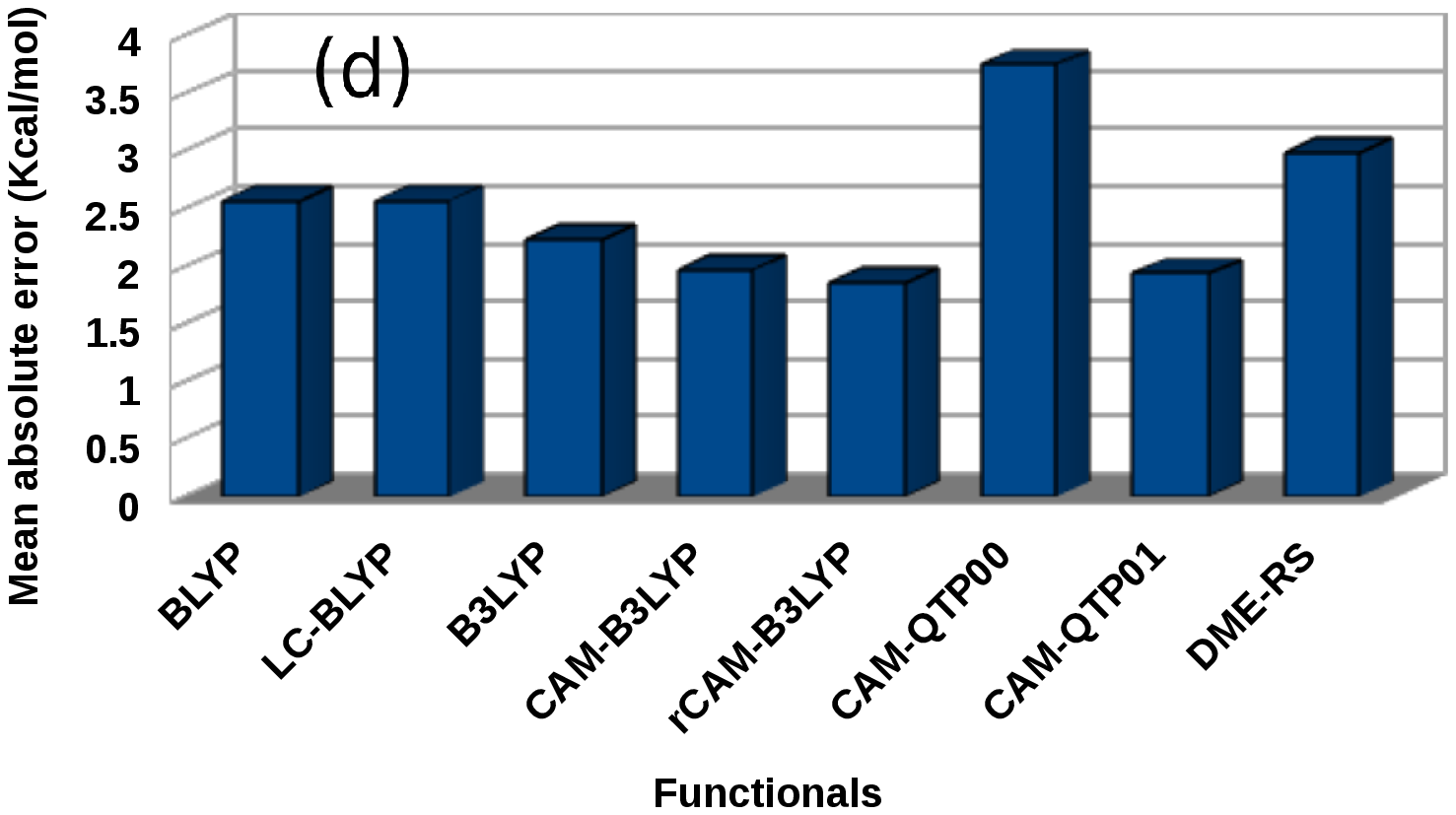}\hfill
\includegraphics[width=.40\linewidth]{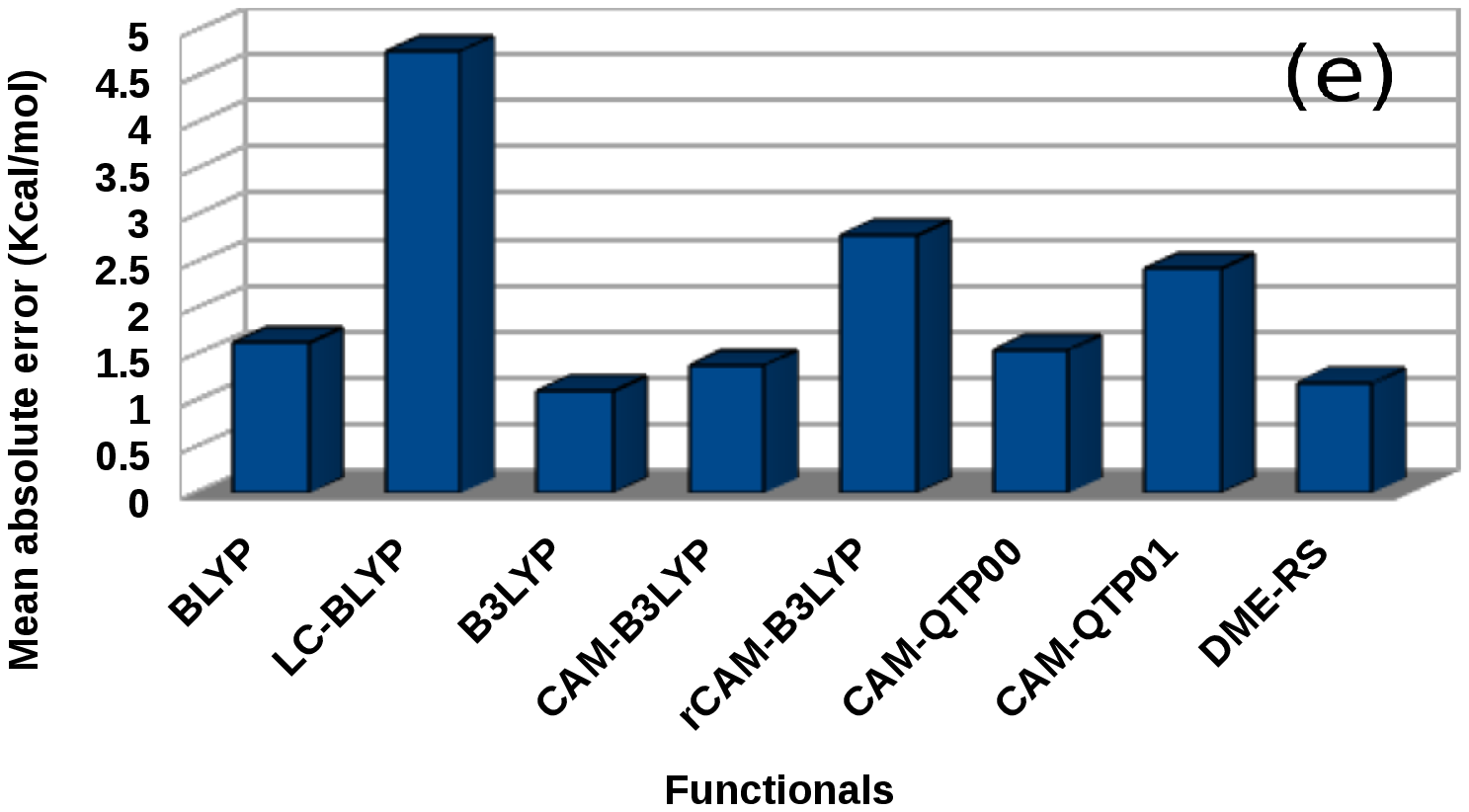}
\includegraphics[width=.40\linewidth]{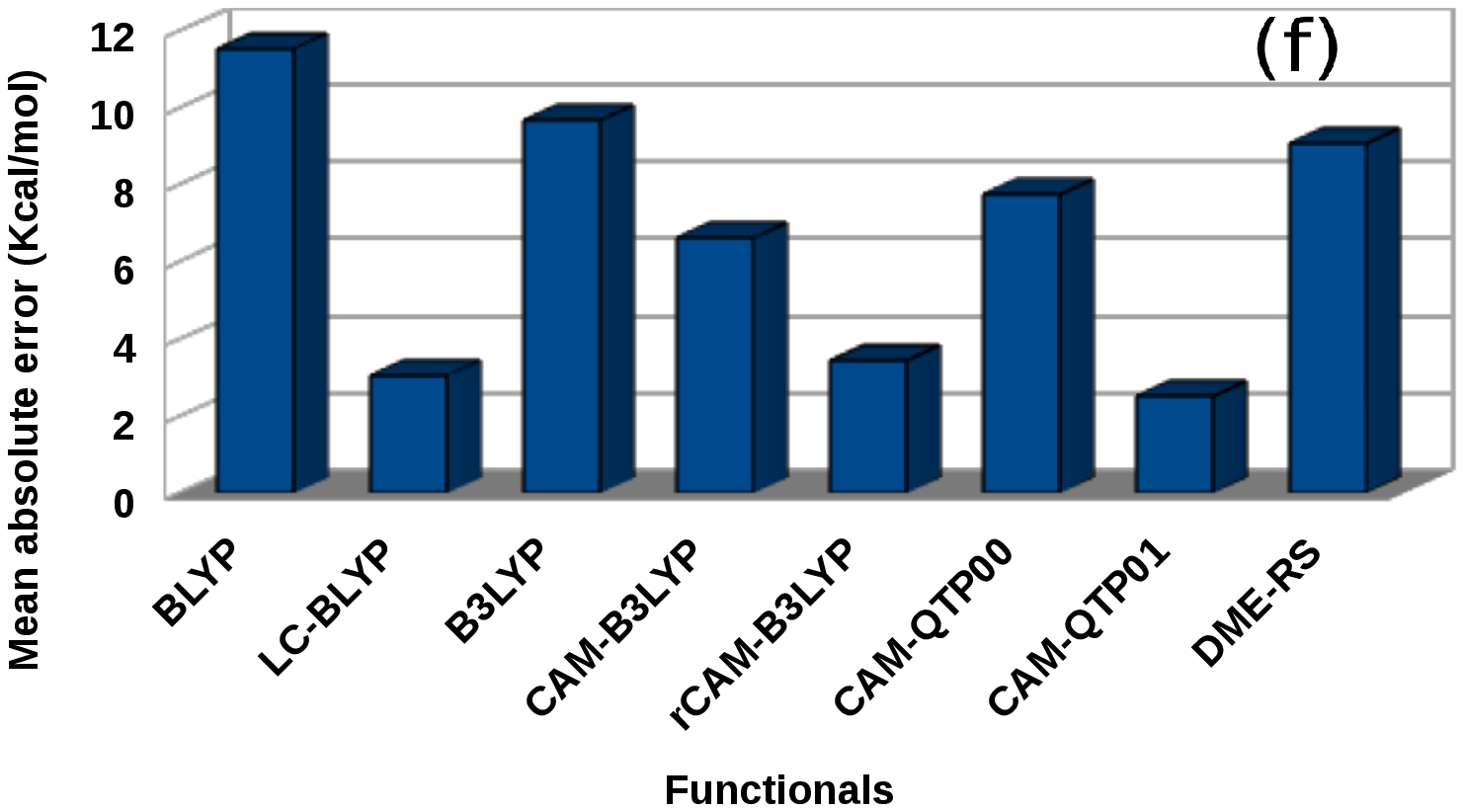}
\includegraphics[width=.40\linewidth]{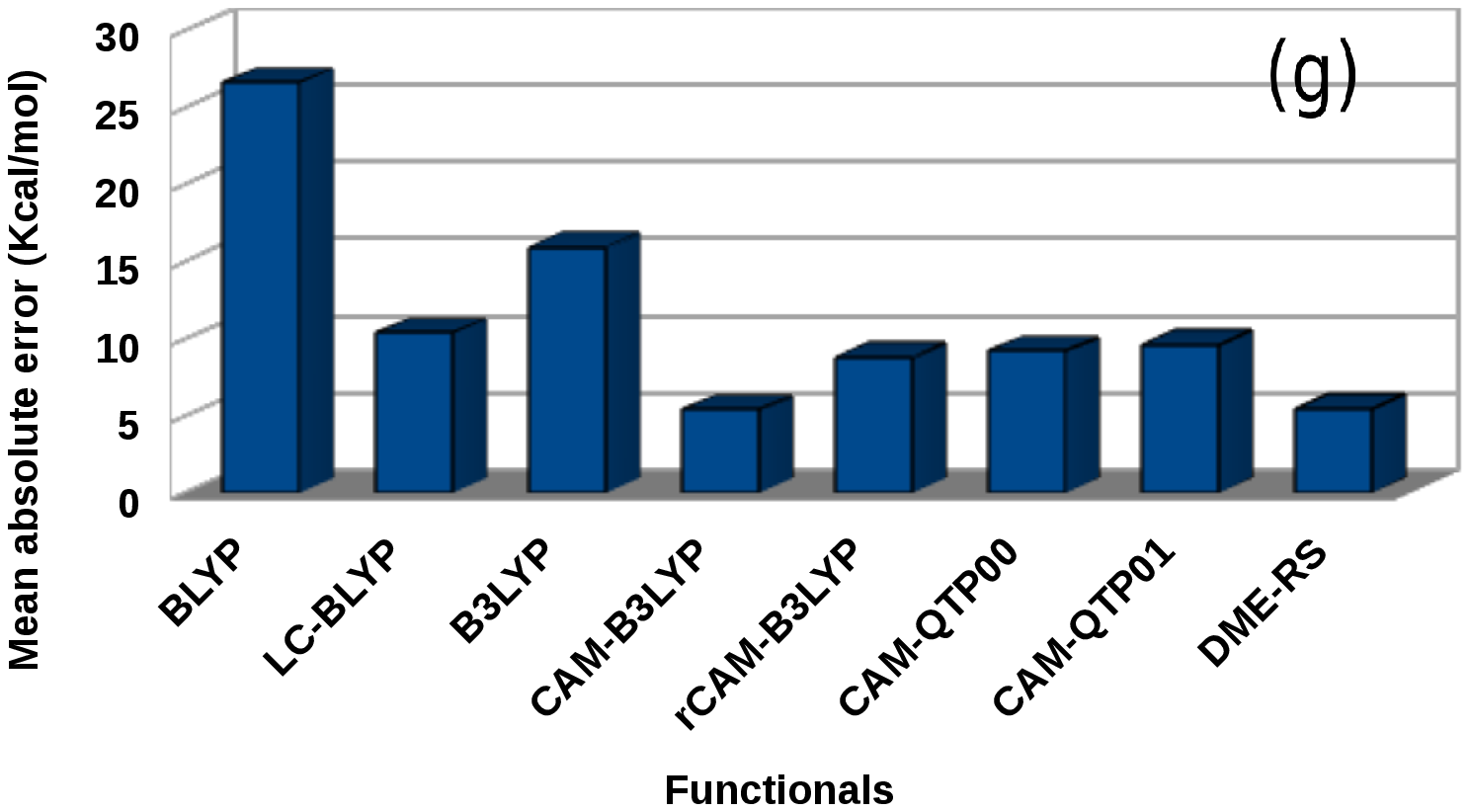}
\includegraphics[width=.40\linewidth]{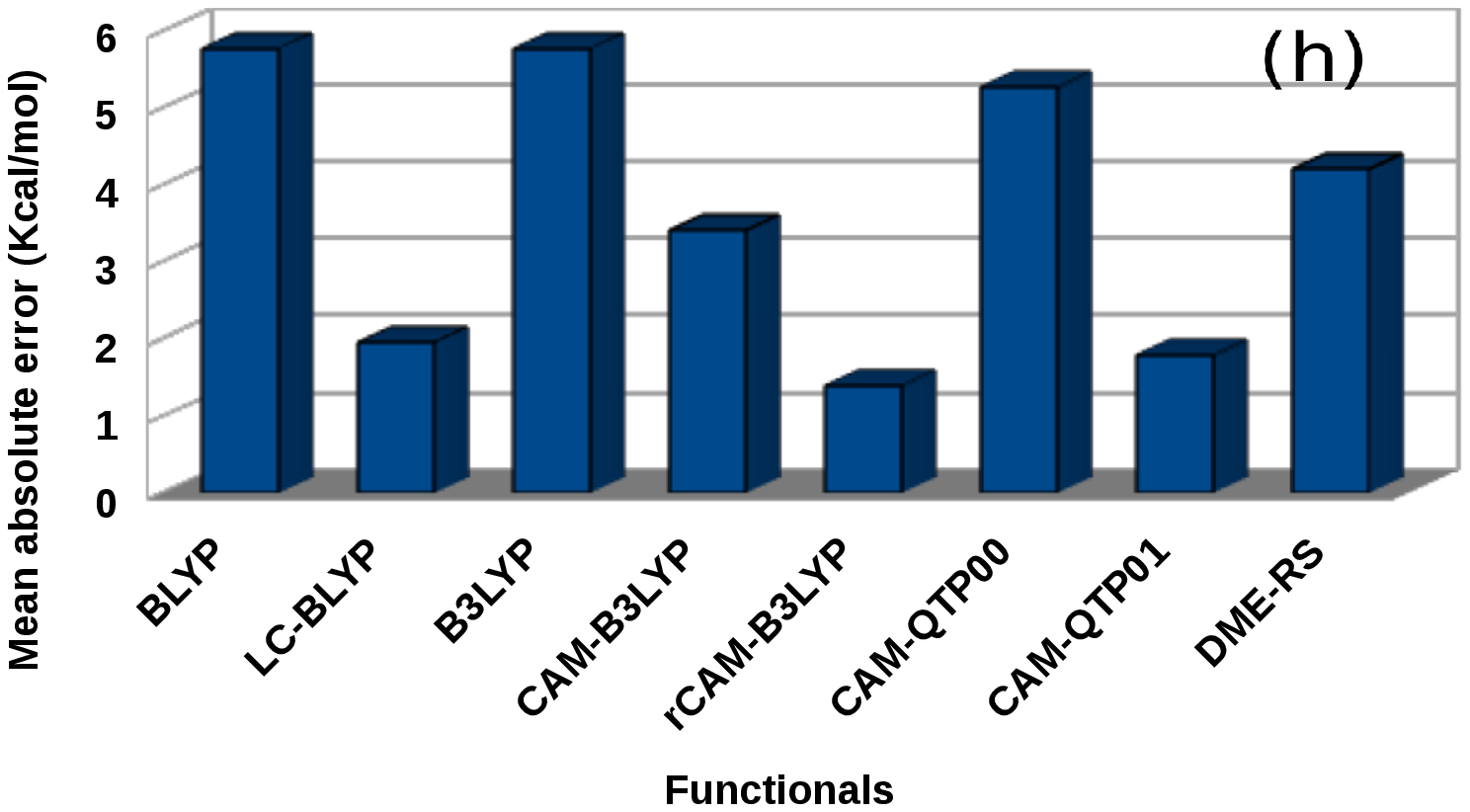}
\includegraphics[width=.40\linewidth]{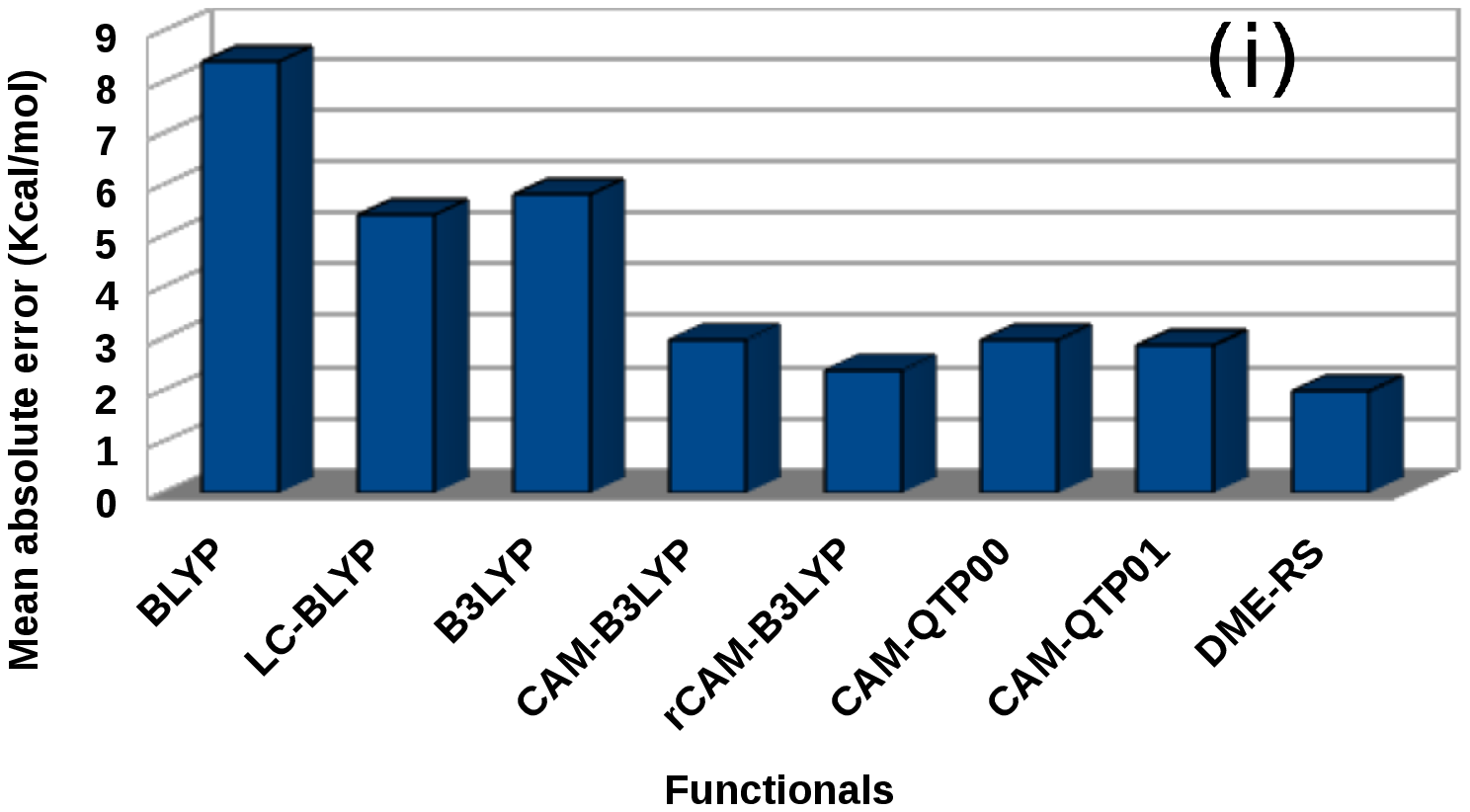}
\includegraphics[width=.40\linewidth]{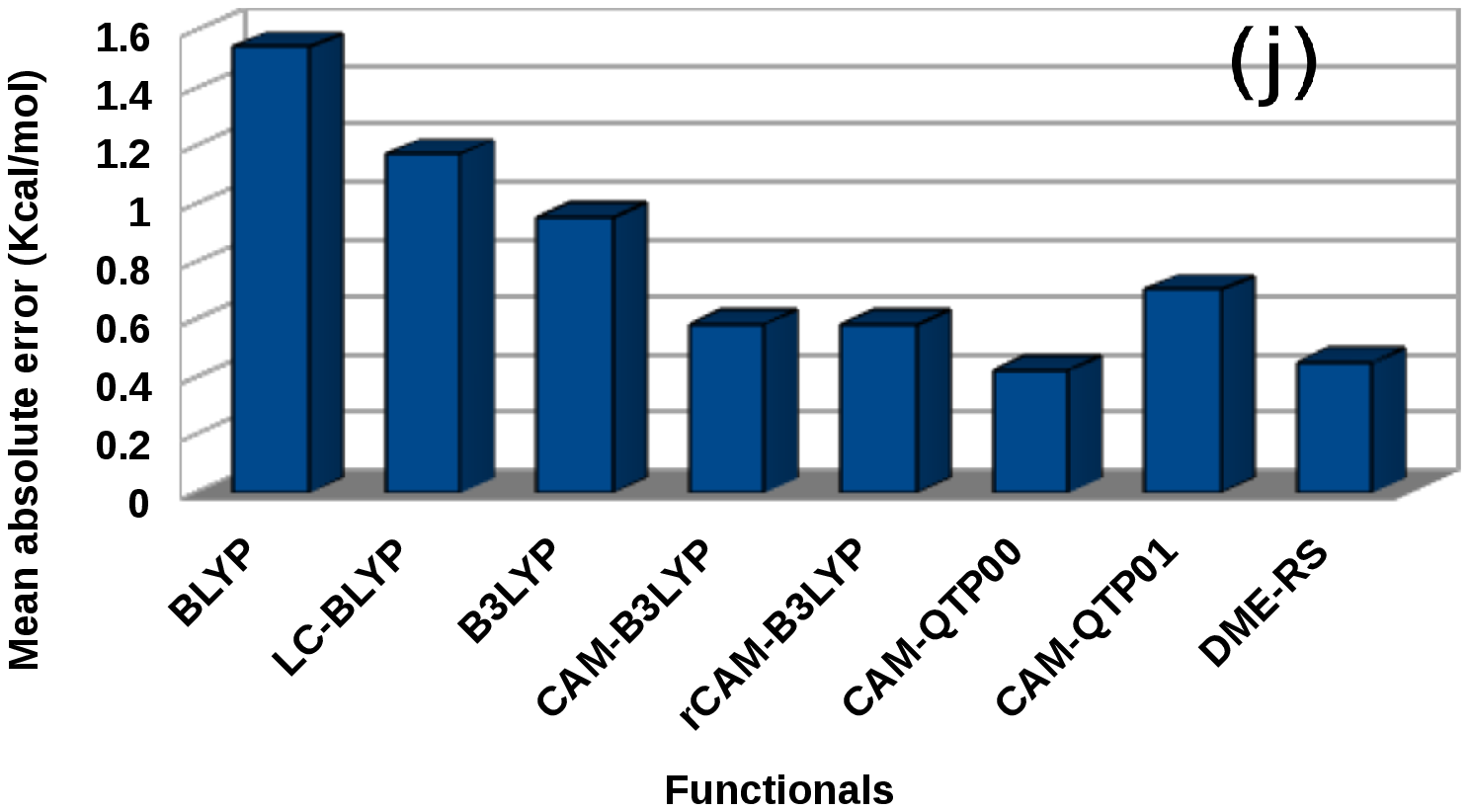}
\caption{Histograms of mean absolute error of different functionals using (a) atomization energy of G2-148 
molecules,  (b) 11 isomerization energies of large molecules, (c) 21 ionization potentials, (d) 13 electron affinities (e) 
8 proton affinities, (f) 12 alkali bond dissociation energies, (g) 11 hydrocarbon chemistry, (h) 13 thermochemistry of 
$\pi$ systems, (i) 76 (non)hydrogen transfers barrier heights, (j) 31 noncovalent complexion energies. 6-311++(3df,3pd) 
basis set is used for all the calculations.}
\label{fig1}
\end{figure*}

\section{Range Seperated Hybrid With Semilocal Exchange Hole}
Now we will propose a meta-GGA level range separated hybrid using recently proposed  semilocal exchange hole~\cite{Tao-Mo16}. 
The exchange hole is localized through the generalized coordinate transformation and normalized by passing inhomogeneity parameter
through Thomas-Fermi wave vector. Using the semilocal form of exchange hole the short range part of the present range 
separated hybrid can be found from the following equation, 
\begin{equation}
 E_x^{SR,DME} = -\frac{1}{2}\int~d^3r\rho(\mathbf{r})\int\frac{1-erf(\mu u)}{u}
 \rho_x^{\rm TM}(\mathbf{r},\mathbf{r}+\mathbf{u})~d^3u,
 \label{eq15}
\end{equation}
where $\rho_x^{\rm TM}(\mathbf{r},\mathbf{r}+\mathbf{u})$ is the DME exchange hole proposed by Tao-Mo~\cite{Tao-Mo16}.
Thus substituting TM DME exchange hole in Eq.(\ref{eq15}), the short range part of semilocal exchange energy becomes
~\cite{patra},
\begin{widetext}
\begin{equation}
\begin{split}
 E_x^{SR,DME}=-\int~\rho\epsilon_x^{unif}\Big[\frac{1}{f^2}\Big\{1-\frac{8}{3}a\Big(\sqrt{\pi}erf(\frac{1}{2a})+(2a-4a^3)
 e^{-\frac{1}{4a^2}}-3a+4a^3\Big)\Big\}+\frac{7\mathcal{L}}{9f^4}\Big\{1+24a^2\Big((20a^2-64a^4)\\
 e^{-\frac{1}{4a^2}}-3-36a^2+64a^4+10\sqrt{\pi}erf(\frac{1}{2a})\Big)\Big\}
  +\frac{245\mathcal{M}}{54f^4}\Big\{1+\frac{8}{7}a\Big((-8a+256a^3-576a^5\\
 +3849a^7-122880a^9)e^{-\frac{1}{4a^2}}+24a^3(-35+224a^2-1440a^4+5120a^6)+2\sqrt{\pi}(-2+60a^2)erf(\frac{1}{2a})
 \Big)\Big\}\Big],
 \label{eq16}
 \end{split}
\end{equation}
\end{widetext}
where $\epsilon_x^{unif}=\frac{9\pi\rho}{4k_{f}^2}$, $f=[1+10(70y/27)+\beta y^2]^{1/10}$, $\mathcal{L}=[3(\lambda^2-\lambda
+1/2)(\tau-\tau^{\rm unif}-|\nabla n|^2/72n)-(\tau-\tau^{unif})+\frac{7}{18}(2\lambda-1)^2\frac{|\nabla\rho|^2}{\rho}]/
\tau^{\rm unif}$, $\mathcal{M}=(2\lambda-1)^2~p$, $a=\frac{\mu}{2fk_f}$, $k_f=(3\pi^2\rho)^{\frac{1}{3}}$, $\tau^{\rm unif}
=\frac{3}{10}k_f^2\rho$, $y=(2\lambda-1)^2~p$ and $p=\frac{|\nabla\rho|^2}{(2k_f\rho)^2}$.

This is the short range part of our current range separated hybrid and for long range we have used HF. In TABLE I we presented 
the range separated parameters and exchange and correlation for different range separated hybrids according to Eq.(\ref{eq10}) 
and Eq.(\ref{eq11}). From TABLE I it is clear that the present range separated functional use only LYP correlation with the DME based 
short range corrected exchange energy. We named our functional as {\bf DME-RS} as it is density matrix expansion based range 
separated functional. 

\begin{table}[!htb]
\label{tableaa}
\caption{Values of different parameters used in range separated functionals}
\begin{tabular}{|c  ||c  c  c  c || c|  c  c  c }
\hline
 & &Exchange& & &Correlation\\ \hline
Name & $\alpha$  &$\beta$ & $\mu$ & $\delta$ & $\gamma$ \\ \hline
CAM-B3LYP    & 0.19 &0.46       &0.33         &0.0           &0.81         \\ 
rCAM-B3LYP    & 0.18352 &0.94979       &0.33         &0.13590           &1.0         \\ 
LC-BLYP    & 0.0 &1.0       &0.33         &0.0           &1.0         \\
CAM-QTP00    & 0.54 &0.37        &0.29         &0.0           &0.80         \\
CAM-QTP01    & 0.23 &0.77       &0.31         &0.0           &0.80         \\
DME-RS    & 0.0 &1.0       &0.33         &0.0           &1.0         \\
\hline
\end{tabular}
\end{table}
Combining all the SR and LR part our DME-RS hybrid exchange correlation functional,
\begin{equation}
 E_{xc}^{DME-RS}=E_x^{SR,DME}[\rho,\nabla\rho,\tau]+E_x^{LR,HF}+E_c^{LYP}
\end{equation}
In our case, we fixed the $\lambda$ and $\beta$ value as it is pescribed in TM functional~\cite{Tao-Mo16}
i.e., $\lambda=0.6866$ and $\beta=79.873$.

\begin{table*}[ht]
\caption{Summary of deviation  using different methods. Mean absolute error (MAE) are calculated here. All MAE are given in 
Kcal/mol unit.}
\begin{tabular}{c  c  c  c  c  c  c  c  c c c c c }
\hline\hline
Method     &BLYP  &LC-BLYP  &B3LYP  &CAM-B3LYP  &rCAM-B3LYP  &CAM-QTP00 &CAM-QTP01  &DME-RS  \\ \hline
G2/148       &10.94      &23.20         &4.31        &4.33&5.41&14.16&9.89&5.09            \\
IsoL6/11 &3.65      &1.82         &2.54        &2.03 &2.27 &2.32 &2.03&1.96                \\
IP21  &4.38      &7.39         &5.29        &4.56 &9.08 &9.30 &8.74 & 3.92               \\
EA13/03  &2.56      &2.56         &2.22        &1.96&1.85&3.74&1.93&2.97                \\
PA8/06     &1.62      &4.77         &1.10        &1.37&2.79&1.54&2.42&1.19                  \\
ABDE12     &11.51      &3.04         &9.67        &6.63&3.41&7.77&2.52&9.08                  \\
HC7/11     &26.56      &10.33         &15.93        &5.35&8.77&9.16&9.58&5.37                  \\
$\pi$TC13     &5.77      &1.95         &5.77        &3.40&1.39&5.28&1.79&4.19                  \\
(N)HTBH76/08     &8.38      &5.39         &5.79        &2.99&2.39&2.97&2.85&1.98                  \\
NCCE31/05     &1.54      &1.17         &0.95        &0.58&0.58&0.42&0.70&0.45                  \\
DC9/12     &27.07      &52.93         &20.81        &9.53&16.98&30.33&23.35&15.70                  \\
\hline\hline
\end{tabular}
\end{table*}

\section{Results and Discussion}
The newly developed DME-RS functional is implemented in version 6.6 of NWChem~\cite{nwchem} code. The basis set used for different calculations are 
indicated with the test set we used in our calculations.

\subsection{Thermochemical properties}
We used Minnesota 2.0 data set~\cite{Quest12} for thermochemical tests. Only for atomization energy we used G2/148  
molecular test set~\cite{g21,g22,g23}, where all the geometries are optimized by MP2 level. More specifically, We heve considered 
following Thermochemical tests: (i) G2/148 - atomization energy of 148 molecules ,  (ii) IsoL6/11 - 11 isomerization energies of large molecules, (iii) 
IP21 - 21 ionization potentials, (iv) EA13/03 - 13 electron affinities, (v) PA8/06 - 8 proton affinities, (vi) ABDE12 - 
12 alkyl bond dissociation energies, (vii) HC7/11 - 11 hydrocarbon chemistry, (viii) $\pi$TC13 - 13 thermochemistry of 
$\pi$ systems, (ix) HTBH38/08 - 38 hydrogen transfers barrier heights, (x) NHTBH38/08 - 38 non-hydrogen transfers barrier 
heights, (xi) NCCE31/05 - 31 noncovalent complexion energies, (xii) DC9/12 - 9 difficult cases. All calculations are 
performed using 6-311++(3df,3pd) basis set. The performance of our DME-RS along with all other functionals for thermochemical 
test set are summarized in TABLE II, where we have given the mean absolute errors (MAE) of each test set.

From TABLE II it is evident that B3LYP and CAM-B3LYP are producing equivalence MAE for atomization energies. It is not surprising because 
both the B3LYP and CAM-B3LYP are parametrized to reduce the MAE of atomization energy. Third best is our DME-RS functional. We
obtain MAE of 5.09 Kcal/mol using our DME-RS functional. The rCAM-B3LYP has greater MAE compared to our DME-RS functional. The QTP functionals 
all have greater MAE than rCAM-B3LYP in atomization point of view. The BLYP and LC-BLYP are far away from the accuracy. Next, we have computed the isomerization energies of 11 large molecules and observed that LC-BLYP has least MAE and next 
best is our DME-RS.  In case of ionization 
potential of 21 test set,  DME-RS is superior to other functional with MAE 3.92 Kcal/mol. For the electron affinities of
13 test set rCAM-B3LYP is superior to other functionals. In this case the MAE of DME-RS is 2.97 Kcal/mol. In case of PA, 
our parametrized DME-RS gives almost equivalent results with that of B3LYP. Among QTP functionals, QTP-00 has less MAE than 
rCAM-B3LYP. For the set of 12 alkyl bond dissociation energies (ABDE), CAM-QTP01 is most accurate one. The MAE of DME-RS is 
almost equivalence to that of B3LYP. In the case of HC7/11, both CAM-B3LYP and DME-RS outperformed other functionals. In the 
case of 13 thermochemistry of $\pi$ systems, rCAM-B3LYP is the best performer. The MAE of our DME-RS is less than that of B3LYP in this case. 
For hydrogen and non-hydrogen reaction barrier benchmarks, our DME-RS outperformed all other range separated, hybrid and semilocal 
functionals. It is well known that B3LYP is not a good candidate for reaction barrier, though the performance of its atomization energy is best 
compared to other functionals. Because errors in the reaction barriers mainly occur due to the problem of 
self-interaction error in the transition state. The DME-RS short range corrected functional combined with LYP reduces the self-interaction error 
compared to B3LYP and CAM-B3LYP (discussed later). Actually, the inclusion of KS-KE density makes our functional very good performer than all other GGA based functionals. That is why meta-GGAs are treated as the most advanced functionals in semilocal level 
also due to its energy ingredients are more accurate than GGAs. For the NCCE31 test set DME-RS and CAM-QTP00 equivalently 
give the same MAE, which is the best among all other functionals. For DC9 test set, DME-RS is the second best performer after 
CAM-B3LYP. 

\begin{figure}
\begin{center}
\includegraphics[width=2.8in,height=2.0in,angle=0.0]{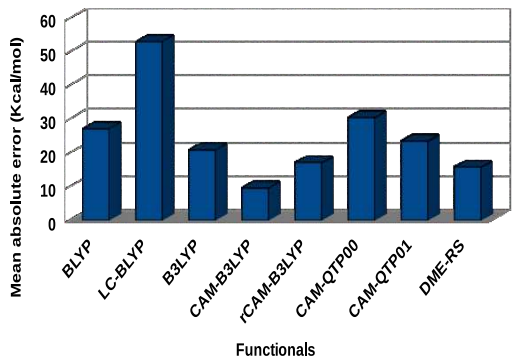} 
\end{center}
\caption{graphical representation of mean absolute error of different functionals using 9 difficult cases of Minessota 2.0
data set. The 6-311++(3df,3pd) basis set is used.}
\label{fig2}
\end{figure}

\subsection{Fraction occupation number in an atom}

\begin{figure*}
\begin{center}
\includegraphics[width=5.8in,height=3.3in,angle=0.0]{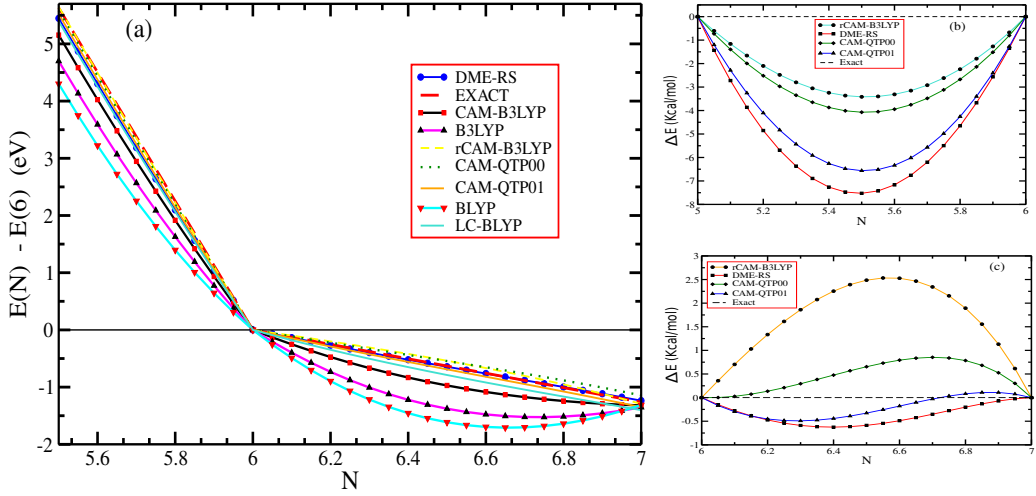} 
\end{center}
\caption{(a) Shown is the energy difference of C atom as a function of electron number. The exact line is obtained using 
experimental -IP and -EA.  Deviation of different functionals from exact piecewise linear behavior are shown for C atom 
for (b) $-1\leq q\leq 0$ and (c) $0\leq q \leq 1$.  The 6-311++(3df,3pd) basis set is used for all the calculations.}
\label{fig1}
\end{figure*}

\begin{figure}
\begin{center}
\includegraphics[width=3.2in,height=2.3in,angle=0.0]{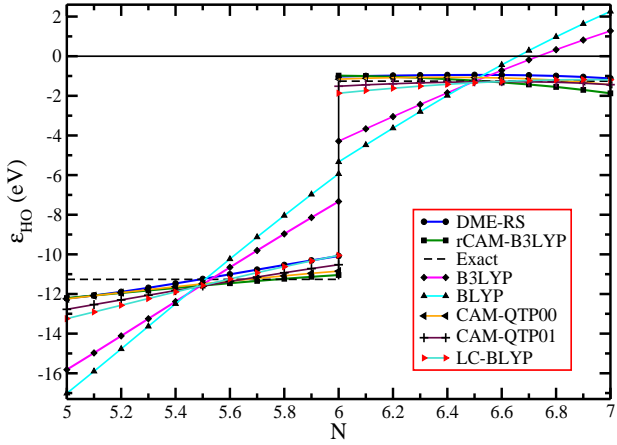} 
\end{center}
\caption{Shown is the highest occupied energy of C atom as a function of electron number N. Exact $\varepsilon_{HO}$ is 
obtained from -IP and -EA. The 6-311++(3df,3pd) basis set is used.}
\label{fig2}
\end{figure}

\begin{table*}
\caption{Comparison of orbital energies of different functionals for C, Li and Ne. All values are in eV. The 6-311++(3df,3pd) 
basis set is used.}
\begin{tabular}{c  c  c  c  c  c  c  c  c }
\hline\hline
Atom&Name  & $-\varepsilon_{LUMO}[N]$  &EA & $-\varepsilon_{HOMO}[N]$ & IP & $-\varepsilon_{HOMO}[N]-[-\varepsilon_{LUMO}[N]]$& 
IP-EA \\ \hline
&BLYP      & 5.005                     &1.642       &5.740         &11.071           &0.735 &9.429         \\
&B3LYP     & 3.972                     &1.727       &7.075         &11.163           &3.103&9.435         \\
&CAM-B3LYP &2.375                      &1.735       &8.824&11.145&6.448&9.410         \\
&rCAM-B3LYP&0.745 &1.729&10.738&11.154&9.992&9.425         \\
C&LC-BLYP  &1.565&1.868&9.751&11.127&8.186&9.258         \\
&CAM-QTP00 &0.934 &1.588&10.489&11.063&9.556&9.475         \\
&CAM-QTP01 &1.240 &1.812&10.194&11.150&8.954&9.338         \\
&DME-RS    &0.794 &1.640&9.925&11.141&9.131&9.501         \\
\hline\hline
&Expt.    & &1.262      &         &11.26           & &10.00         \\
\hline\hline
&BLYP      &1.536 &0.455       &3.081         &5.528           &1.544&5.073         \\
&B3LYP     &1.434 &0.556       &3.648         &5.626           &2.214&5.069         \\
&CAM-B3LYP &1.035 &0.507       &4.647         &5.600           &3.611&5.093         \\
&rCAM-B3LYP&0.538 &0.475       &5.750         &5.619           &5.212&5.144         \\
Li&LC-BLYP &0.593 &0.452       &5.385         &5.581           &4.791&5.129         \\
&CAM-QTP00 &0.828 &0.507       &5.268         &5.591           &4.440&5.084         \\
&CAM-QTP01 &0.780 &0.516       &5.462         &5.621           &4.681&5.105         \\
&DME-RS    &0.564 &0.438       &5.369         &5.522           &4.805&5.084         \\
\hline\hline
&Expt.    &  &0.62        &         &5.39           & &4.77         \\
\hline\hline
&BLYP      &-3.058 &-6.214&13.387&21.702&16.446&27.916         \\
&B3LYP     &-3.704 &-6.197&15.674&21.746&19.378&27.943         \\
&CAM-B3LYP &-4.930 &-6.273&17.693&21.765&22.623&28.039         \\
&rCAM-B3LYP&-6.137 &-6.314&19.906&21.788&26.044&28.103         \\
Ne&LC-BLYP &-5.720 &-6.251&18.160&21.952&23.880&28.203         \\
&CAM-QTP00 &-5.795 &-6.429&20.850&21.463&26.646&27.893         \\
&CAM-QTP01 &-5.834 &-6.288&19.412&21.799&25.247&28.088         \\
&DME-RS    &-6.126 &-6.457&18.163&21.745&24.290&28.203         \\
\hline\hline
&Expt.     &       &-5.50 &      &21.56 &      &27.06   \\
\hline\hline
\end{tabular}
\end{table*}
Studying fractional occupation number in an atom is important from the different physical point of view as it is directly related 
to the solid state properties involving band gap. There are several literature that have considered this phenomena~\cite{vsp07,
mcy06,rpcvs06,pz81,prp14,hkkk12,phl84,rpcvs07,pplb82,rmjb17}. Here, we consider the C atom to demonstrate the deviation of 
range separated functionals from exact straight line behavior. 

In Fig-(\ref{fig1}) we demonstrated the behavior of our functional along with all other B88 family functionals. The C atom is neutral 
at atomic number 6. We varied total particle number from 5 to 7 in steps 0.05 ($q$). In between $5\leq N<6$ 
the exact behavior is obtained from the experimental ionization potential (IP) data, whereas, in between $6<N\leq7$ it is obtained from 
the experimental electron affinity (EA). Thus the exact straight line behaviors are just experimental IP and EA. The only semilocal function 
we have used for comparison is BLYP functional. Semilocal functional has inherent delocalization error, which delocalized both HOMO and LUMO  
from its exact behavior. HF has a tendency to overlocalize the HOMO and LUMO. Therefore, the inclusion of HF exchange with semilocal form actually compensates 
both the localization and delocalization error. Most impressive error minimization is observed for rCAM-B3LYP, DME-RS, LC-BLYP, 
CAM-QTP00 and CAM-QTP01. The LC-BLYP is long range corrected version of BLYP. As expected, it minimizes the delocalization error due 
to the inclusion of HF exchange. The CAM-BLYP also shows significant delocalization error but less than B3LYP. The CAM-B3LYP shows more 
delocalization in between $6\leq N\leq 7$. In this region larger delocalization means unstable C$^-$ anion. The rCAM-B3LYP improves 
upon CAM-B3LYP because it has been parametrized to reduce the error of delocalization of CAM-BLYP. In C$^+$ (cation) rCAM-B3LYP, CAM-QTP00, and CAM-QTP01 match exactly with the exact straight line. In this case our meta-
GGA range separated DME functional slightly deviates from rCAM-B3LYP, CAM-QTP00, and CAM-QTP01. Interestingly, we obtain almost exact 
straight line behavior of DME-RS in the range of $6<N\leq7$, where slight over localization is observed for rCAM-B3LYP and CAM-QTP00 
and slight delocalization observed in case of CAM-QTP01. Our meta-GGA DME-RS dme functional therefore superior in case of 
describing the stability of C$^-$ anion. In Fig-(\ref{fig1}b) and Fig-(\ref{fig1}c), we have shown the deviation of different 
functionals from exact piecewise linear behavior for C atom in the range $-1\leq q\leq 0$ and $0\leq q \leq 1$. Here, we have  
calculated $\Delta E$ by using following formula:
\begin{eqnarray}
 \Delta E_{q\exists[0,1]} &=& E(N_0+q)-[(1-q)E(N_0)+qE(N_0+1)]\nonumber\\ 
 \Delta E_{q\exists[-1,0]} &=& E(N_0+q)-[(1+q)E(N_0)-qE(N_0-1)]\nonumber\\
\end{eqnarray}
Comparison are shown among those functionals which are close to exact one. From Fig-(\ref{fig1}b) and Fig-(\ref{fig1}c), it is clear 
that in $-1\leq q\leq 0$ rCAM-B3LYP closer to the exact one and in $0\leq q \leq 1$ rCAM-B3LYP and CAM-QTP00 has a tendency to 
slightly over delocalize, whereas CAM-QTP01 and DME-RS are more close to exact one. The stability of C$^+$ cations and 
C$^-$ anions also reflects in Fig-(\ref{fig2}), where we have shown the energy of highest occupied energy by changing the occupation 
number. The BLYP and B3LYP show positive $\varepsilon_{HO}$ value, which is an artifact of the instability of C$^-$ anion for 
those functionals. For $5\leq N \leq 6 $ all the hybrid functionals deviate more or less from exact linearity of -IP. In $6 
\leq N \leq 7 $, DME-RS shows almost linear dependence of total energy on N, predicts accurate total energy of C atoms and its 
anions. The CAM-QTP00 and CAM-QTP01 functionals show similar behavior as obtained by rCAM-B3LYP. Generally speaking, our meta-GGA range 
separated hybrid predicts very well the orbital energy and almost shows linear behavior, though it is not designed to optimize 
the error of delocalization. The semilocal exchange hole used here has unique property that was previously missed in rest of the 
range separated hybrid. Therefore, though it is not designed for reducing the delocalization error, it shows some inherent 
localization due to the fact that its obtain from the localized exchange hole.

Now, restoring back to the Janak's theorem~\cite{janak}, which tells us that variation of the total energy w.r.t the 
occupation is equal to the eigenvalue of that orbital. From this theorem, it can be proved that the variation of total 
energy w.r.t the total number of particles is equal to the HOMO eigenvalue as we consider variation of the particle 
number of frontier orbital only, which is either HOMO or LUMO orbital. Therefore, during the occupation number 
variation of frontier orbital,
\begin{eqnarray}
 E(N)-E(N-q)&=&\int_{0}^q~\varepsilon_{HOMO} \partial q\nonumber\\
 {\textit{i.e.,}}~~~~ \lim_{q\to 0}\frac{\partial E(N+q)}{\partial q}&=&\varepsilon_{HOMO}~~-1\leq q< 0.
\end{eqnarray}
or,
\begin{eqnarray}
E(N)-E(N+q)&=&\int_{0}^q~\varepsilon_{LUMO} \partial q\nonumber\\
{\textit{i.e.,}}~~~~ \lim_{q\to 0}\frac{\partial E(N+q)}{\partial q}&=&\varepsilon_{LUMO}~~0< q\leq 1.
\end{eqnarray}
Hence, during the variation of the occupation, energy should fix at  $\varepsilon_{HOMO}$ for $-1\leq q< 0$ and $\varepsilon_{LUMO}$ 
for $0< q\leq 1$. In Fig-(\ref{fig2}), we compared the performance of our DME-RS with all the semilocal, hybrids and range 
separated hybrids constructed from B88 family for C atom. In the range $-1\leq q< 0$ all the hybrid functionals deviate less or more from 
exact linearity of -IP. Whereas, in the range of $0< q\leq 1$ DME-RS shows almost linear dependence of total energy on N. The CAM-QTP00 and CAM-QTP01 
functionals show similar behavior as obtained by rCAM-B3LYP. For better physical insight we compared the HOMO and LUMO energy values 
with IP and EA obtained from different functionals. In TABLE III, we compared the performance of different functionals for C, 
Li and Ne atoms. 

As it is known from exact DFT the highest occupied and lowest unoccupied KS energy eigenvalue are equal to the -IP and -EA. We 
have tested this fact for all the B88 family functionals along with our meta-GGA range separated functionals. Among all functionals, 
CAM-QTP01 quite closely obtain $\varepsilon_{LUMO} = $-EA~(experimental). The CAM-QTP00 is the next functional which obtain the 
LUMO eigenvalue close to the experimental -EA. The rCAM-B3LYP and DME-RS produce almost equivalent LUMO eigenvalue energy. Whereas, 
all other functional overestimated the LUMO. In case of HOMO eigenvalue and -IP, we obtained good agreement with rCAM-B3LYP, which is 
also evident from the fractional occupation curve of Fig-(\ref{fig1}). Our DME-RS obtained almost equivalent results with CAM-QTP01 
and CAM-QTP00 and behave similarly with rCAM-B3LYP. Though the DME-RS is not parametrized by satisfying ionization potential theorem 
or minimizing the error during fractional occupation number. But, it predicts the KS eigenvalues with -IP and -EA quite accurately.
\begin{figure*}
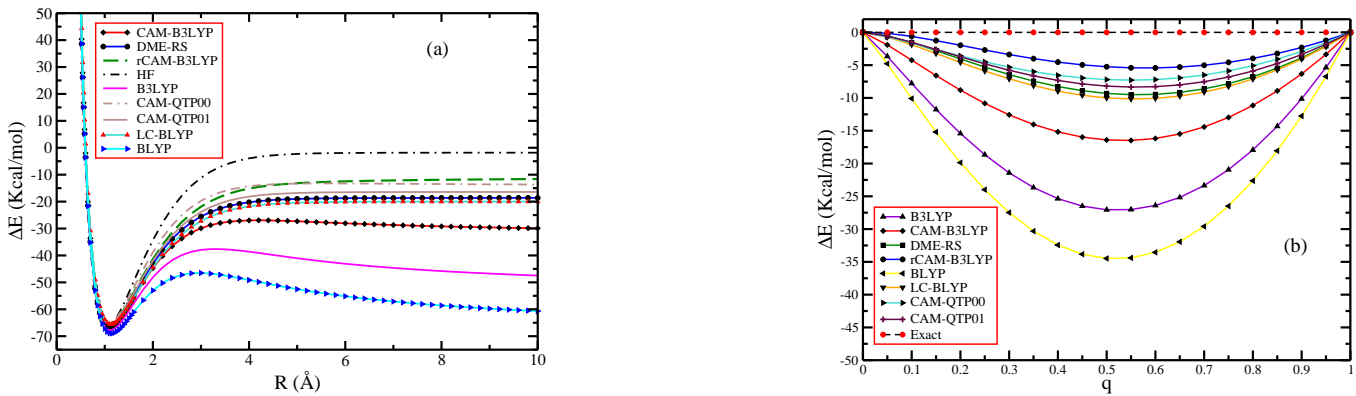

\includegraphics[width=.40\linewidth]{cam-dme.eps}\hfill
\includegraphics[width=.40\linewidth]{h-fc.eps}
\caption{(a) dissociation curve of H$_2^+$ obtained from different functional form and (b) deviation of exact energy of H atom 
for fractional occupation number. The 6-311++(3df,3pd) basis set is used.}
\label{fig3}
\end{figure*}
As we know that getting accurately the HOMO-LUMO energy difference is important in the sense that different excited state properties in the molecular systems are directly 
connected to the orbital energy. Also, solid state band gap is connected to the orbital energies as $-\varepsilon_{HOMO} 
+ \varepsilon_{LUMO}=IP-EA=E_g $ (fundamental band gap). Therefore, it is important to obtain HOMO-LUMO gap accurately. 
It is evident from the analysis of C, Li and Ne results that rCAM-B3LYP determines well with the HOMO-LUMO energy which is 
very close to the experimental IP-EA, because it is parametrized to reduce the error of the fractional occupation number. Next 
good comparison is found for CAM-QTP00. Both QTP functionals are designed by satisfying Berlett's IP theorem. 
In the case of Li, we found that DME-RS outperformed rCAM-B3LYP and CAM-QTP functionals in predicting HOMO-LUMO energy difference 
and its equivalence with the experimental IP-EA. Whereas, in the case of Ne, rCAM-B3LYP, CAM-QTP and DME-RS perform almost equivalently.

\subsection{Dissociation curve of H$_2^+$}
Another challenge in quantum chemistry is to predict accurately the dissociation curve of H$_2^+$ molecule, which is a paradigm 
system in quantum chemistry. Semilocal DFA failed to describe the dissociation curve properly. The fundamentally wrong 
prediction of dissociation curve is due to the inherent limitation of delocalization error in case of fractional occupation 
number. Perdew-Zunger self-interaction correction improves the dissociation curve as shown in ref.~\cite{vsp07,rpcvs06,rpcvs07}. 
For H$_2^+$ molecule, HF is taken to be exact one because no correlation present in the system. At $R=10 \AA{}$ each H atom has 
fractional (0.5) electron, which is beyond the limit of semilocal functional to describe accurately. The semilocal BLYP functional 
deviates from the exact value more as shown in Fig-(\ref{fig3}). B3LYP only improves moderately the dissociation curve over BLYP 
and predict wrong dissociation limit. Though CAM-B3LYP improves the dissociation curve, still it is almost 46 Kcal/mol less than 
HF. Most significant improvements are observed for rCAM-B3LYP, CAM-QTP00, CAM-QTP01, LC-BLYP and our proposed DME-RS. On the 
right side of Fig-(\ref{fig3}), we obtain the deviation of the exact energy of H atom for fractional occupation number. Here, we 
have plotted the energy difference computed at the integer particle numbers to that of the piecewise linear interpolation i.e.,
\begin{equation}
 \Delta E = E(N+q)-[(1-q)E(N)+q E(N+1)]
\end{equation}
It is obtained that parametrized rCAM-B3LYP performed well among QTP and DME-RS. The deviation of rCAM-B3LYP, QTP-00, QTP-01 and 
DME-RS at $q=\frac{1}{2}$ are -5.42 Kcal/mol,-7.15 Kcal/mol, -8.43 Kcal/mol and -9.52 Kcal/mol respectively.

It is also noteworthy to mention that at one particle framework, the energy of H atom should equal to the HOMO energy value.
HF, which is exact in one particle framework only satisfy this condition. All other functionals seem to deviate from this 
exact conditions. 
\begin{table}[h]
\caption{Energy and HOMO eigenvalue of hydrogen atom obtain from different functionals. All values are in Hartree. The 6-311++
(3df,3pd) basis set is used.}
\begin{tabular}{c  c  c  c  c  c  c  c  c }
\hline\hline
          &Energy(Ha)  &~~~~~~~~HOMO(Ha) \\ \hline
HF        &-0.499        &~~~~~~-0.499       \\
BLYP      &-0.497        &~~~~~~-0.272       \\
B3LYP     &-0.502        &~~~~~~-0.322        \\
CAM-B3LYP &-0.498        &~~~~~~-0.386         \\
rCAM-B3LYP&-0.496        &~~~~~~-0.455         \\
LC-BLYP   &-0.490        &~~~~~~-0.417         \\
CAM-QTP00 &-0.499        &~~~~~~-0.475         \\
CAM-QTP01 &-0.496        &~~~~~~-0.435         \\
DME-RS    &-0.498        &~~~~~~-0.434         \\
\hline\hline
\end{tabular}
\end{table}

\subsection{Fraction occupation number on dissociation limit}
\begin{figure}[h]
\begin{center}
\includegraphics[width=3.0in,height=2.0in,angle=0.0]{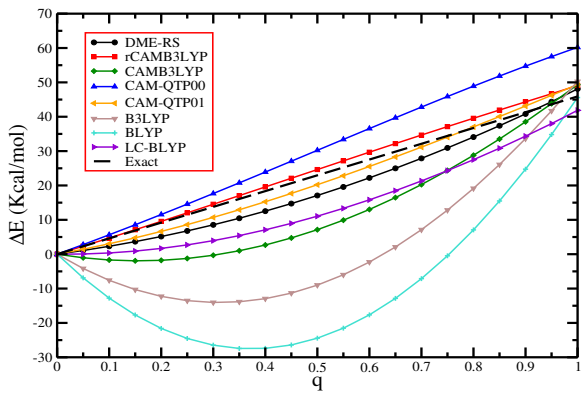} 
\end{center}
\caption{The energy difference of the dissociation of LiF shown using different functionals with respect to fractional charges
through the reaction Li + F$\to$ Li$^+$+F$^-$. The 6-311++(3df,3pd) basis set is used.}
\label{fig4}
\end{figure}

LiF molecule seems to be a good example to discuss the influence of the fractional change during dissociation of a molecule.
This example has been considered in previous work also~\cite{vsp07,rmjb17}. Here, we will compare our DME-RS with other
range separated hybrids, hybrids and semilocal functionals. Since, IP(Li)$>$EA(F), it dissociates into neutral Li and F atom. It has 
been shown that in molecular dissociation limit DFAs show spurious behavior and deviates from the exact straight line. To 
test the performance and behavior of each functional we consider that during dissociation electron flows from Li to F atom, 
make Li$\to$Li$^{+q}$ and F$\to$F$^{-q}$. Overall, one can write as $Li+F\to Li^{+q}+F^{-q}$ and the difference between the 
energy of $Li^{+q}+F^{-q}-(Li+F)=\Delta E $ is the measure of the performance of the different functionals. Here, we have calculated 
the total energy of separate subsystem with the fractional number different from that of a neutral total system. the performance 
of all the functional is presented in Fig-(\ref{fig4}). The semilocal functionals are found to be accurately predict the total 
energy for the integer charge as shown for BLYP in Fig-(\ref{fig4}). Both $q=0$ and $q=1$, BLYP matches perfectly well with that 
of exact one. But for fractional occupation number it deviates quite largly than all other hybrid and range separated functionals. 
This is due to the fact that semilocal functional has an inherent problem of MESI during fractional charge particle. Semilocal 
functionals have a tendency to predict accurate energy at in integer particle number but not in between two integer numbers.  
Semilocal functionals are designed to satisfy the total energy at integer particle number. B3LYP seems to be better than BLYP 
because it mixes HF exchange, but still, it shows minimum nature of curve instead of the actual straight line. Now, we consider 
the comparison among all range separated functionals. Among all range separated functionals rCAM-B3LYP very close to the actual 
straight line. After CAM-QTP01 functional and DME-RS seems to be very close to the actual one. The DME-RS has slight minimum nature as
shown in  Fig.(\ref{fig4}). CAM-QTP00 match exactly at $q=0$, but deviates almost 13 Kcal/mol at $q=1$.
CAM-B3LYP show deviation of almost $2$ Kcal/mol at $q=1$. Our DME-RS is very close at $q=1$ than all other functionals.

\subsection{B3LYP hybrid versus other range separated hybrid}
The role of exact exchange was first proposed by Becke in the series of seminal papers~\cite{Becke14}. The first step towards 
the succesful attempt to mix globally HF exchange with semilocal functional came with B3LYP hybrid, which is still all in one very good performer for the 
quantum chemist. B3LYP is a GGA type functional, which uses only density and gradient of density. Later, it has been realized 
that in spite of its grand success for thermochemistry there are few cases in which B3LYP failed to achieve accuracy. Especially, 
in describing the reaction barrier heights and solid state properties where fractional occupation number play important role. It has been observed that B3LYP 
is far from accuracy in describing reactions barriers. As a source of errors, it has been realized that semilocal functional and 
hybrid B3LYP deviates from the actual straight line nature of fractional occupation number. Beyond the dawn of B3LYP, the 1st 
range separated hybrid using the B88 exchange was proposed by Tozer et. al.~\cite{camb3lyp}. Their CAM-B3LYP used HF in the long 
range part and B88 in short range using the range separation. It has been observed that in many cases CAM-B3LYP outperformed 
B3LYP without hindering accuracy of atomization energies. Interestingly, in describing H$_2^+$ dissociation curve CAM-B3LYP is 
far better than B3LYP. It automatically reduces the error of reaction barriers also. Later, Cohen et. al.~\cite{mcy06} further 
parametrized CAM-B3LYP by reducing the MESI error by realizing that MESI solves many problems including dissociation limit and 
reaction barriers. Therefore, the search for better functional than B3LYP hybrid continuous to be an active research field 
with promisingly new perspective. Further modifications of CAM-B3LYP was done by Ranasinghe et.al.~\cite{rmjb17} in their QTP 
functionals. They used Barlett IP theorem to fix the parameters related to the the range separation functionals. So long all 
the proposed modifications were within GGA formalism. Beyond GGA formalism, we here proposed a meta-GGA type range separated 
functional using recently proposed DME exchange hole by Tao-Mo~\cite{Tao-Mo16}. The range separation parameters we used here 
was fixed to reduces the MAE of atomic total energies, this automatically reduces the error of atomization energies over rCAM
-B3LYP and QTP functional. Though our functional is not designed to reduce the MESI, it has been observed that it is a very 
good performer in many cases, especially in describing reaction barriers height. The dissociation curve we obtain is much better 
than CAM-B3LYP. The main reason of the improvement can be explained by the inclusion of KS-KE. It is well-known fact that most 
advanced semilocal functional are meta-GGAs. Therefore, no doubt that using meta-GGA exchange hole also improve all the 
properties that are achievable using range separated hybrids.

\section{Conclusions}
We have performed several comprehensive testing of our KS-KE dependent DME based meta-GGA type long range corrected range 
separated hybrid with all B88 family based semilocal, hybrid and range separated hybrid. For comparison we choose only the 
B88 family functionals because they use LYP and VWN correlation. LYP correlation we used with our DME-RS functionals. We 
have tested our functional from thermochemical properties using Minessota 2.0 data set to all the properties related to the 
fractional occupation number. BLYP, LC-BLYP, B3LYP and CAM-B3LYP are functionals that have been designed to less the errors 
of atomization energies. It has been shown that our DME based range separated functional is a very good performer for IsoL6, 
PA8, HC7, barrier height and NCCE31 data set. Not only that, for properties related to the fractional occupation number our 
DME-RS better than B3LYP and CAM-B3LYP, though it is not designed to reduce the MESI error. Other functionals like rCAM-B3LYP 
is a very good performer because it is parametrized to reduces to MESI and QTP functionals are parametrized to satisfy Berlett's 
IP theorem. The main reason for the improvement of our DME-RS over other range separated functionals are because it is constructed 
using  full spherical averaged normalized meta-GGA type exchange hole. The exchange hole also used generalized coordinate 
transformation, which makes the exchange hole localized. However, the coordinate transformation parameter has been fixed using 
H atom. It is designed using the balanced treatment of localized and conventional exchange hole~\cite{Tao-Mo16}. Lastly, we 
conclude that our DME-RS can be further improved using the method of reduction of MESI, as proposed by Cohen et. al.~\cite{mcy06}. 
Also, using meta-GGA correlation proposed by Tao et. al.~\cite{TPSS03} DME-RS can be further improved. In that case, we have to 
take care of the full Tao-Mo exchange hole~\cite{Tao-Mo16}. 


\end{document}